# Enhancing Operation of a Sewage Pumping Station for Inter Catchment Wastewater Transfer by Using Deep Learning and Hydraulic Model

*Duo Zhang[1]; Erlend Skullestad Hølland[1]; Geir Lindholm[2]; Harsha Ratnaweera[1]*

1. Faculty of Sciences and Technology, Norwegian University of Life Sciences, 1432, Ås, Norway
2. Rosim AS, Brobekkveien 80, 0582, Oslo, Norway

**Abstract:** This paper presents a novel Inter Catchment Wastewater Transfer (ICWT) method for mitigating sewer overflow. The ICWT aims at balancing the spatial mismatch of sewer flow and treatment capacity of Wastewater Treatment Plant (WWTP), through collaborative operation of sewer system facilities. Using a hydraulic model, the effectiveness of ICWT is investigated in a sewer system in Drammen, Norway. Concerning the whole system performance, we found that the Søren Lemmich pump station plays a vital role in the ICWT framework. To enhance the operation of this pump station, it is imperative to construct a multi-step ahead water level prediction model. Hence, one of the most promising artificial intelligence techniques, Long Short Term Memory (LSTM), is employed to undertake this task. Experiments demonstrated that LSTM is superior to Gated Recurrent Unit (GRU), Recurrent Neural Network (RNN), Feed-forward Neural Network (FFNN) and Support Vector Regression (SVR).

**Author names and affiliations:**

Duo Zhang (corresponding author):

   Ph.D. candidate, Faculty of Science and Technology, Norwegian University of Life Sciences

   Email: Duo.Zhang@nmbu.no

Erlend Skullestad Hølland

   Master student, Faculty of Science and Technology, Norwegian University of Life Sciences

Geir Lindholm:

   CEO, Rosim AS, Brobekkveien 80, 0582, Oslo, Norway

   Email: geir@rosim.no

Harsha Ratnaweera:

   Professor, Faculty of Science and Technology, Norwegian University of Life Sciences

   Email: Harsha.Ratnaweera@nmbu.no

# 1. Introduction

Control overflow from the sewer system and Wastewater Treatment Plant (WWTP) is a crucial and challenging task for many cities in developed countries, such as the Drammen city in Norway. The Drammen city is located in southeastern Norway, it has two wastewater treatment plants: the Muusøya WWTP and the Solumstrand WWTP. The Muusøya WWTP has a designed treatment capacity of 33,000 PE (population equivalents), a dimensioning flow (Qdim) of 780 m$^3$/h, and a maximum flow (Qmax) of 1,200 m$^3$/h. The Solumstrand WWTP has a designed treatment capacity of 130,000 PE, the Qdim and Qmax for the Solumstrand WWTP is 2,000 m$^3$/h and 4,000 m$^3$/h respectively. Combined sewer accounts for more than 80% and less than 50% respectively in the sewer system associated with the Muusøya WWTP and the Solumstrand WWTP. Moreover, the drainage area of the Muusøya WWTP has a higher population density than the rest of Drammen due to it is located in the traditional city center. The lower WWTP capacity, a higher portion of combined sewer, and denser population have resulted in the severe overflow problem in the Muusøya area. Therefore, the Drammen city launched the Regnbyge 3M project to mitigate overflow from the sewer system and the WWTP of Drammen.

There are two types of overflow mitigation measures: structural measures and nonstructural measures (Lee et al., 2017). Structural measures refer constructing new hydraulic facilities and the rehabilitation of sewer components (e.g., expansion of sewer pipes). Nonstructural measures are methods that maximize the capacity of the sewer system with minimal changes to the infrastructure through intelligent operating strategies. As the most popular structural measures, the storage tank is still servicing in many developed cities. However, due to limited space or high investments, storage tanks cannot be always constructed in densely populated urban context (Ganora et al., 2017; Ngo et al., 2016) such as the Muusøya area. The drawbacks of structural measures have motivated the research for nonstructural methods, such as exploit the sewer in-line storage capacity (Darsono and Labadie, 2007; Grum et al., 2011; Garofalo et al., 2017), intelligent sewer control (Lee et al., 2017) and explore underground space (Wu et al., 2016).

Considering the spatial mismatch of capacity of the sewer system and WWTP between the Muusøya

WWTP and the Solumstrand WWTP, as well as according to the goal of the Regnbyge 3M project, we propose a novel nonstructural method: Inter Catchment Wastewater Transfer (ICWT) for the Drammen city. The idea of ICWT is inspired by the concept of Inter-basin Water Transfer (IBWT). IBWT refers to transfer water from basins having sufficient water (donor basin) to basins facing water shortages (receiving basin) (Wang et al., 2015, Yevjevich 2001). IBWT utilizes the differences of flow regime in different basins to create a win-win situation.

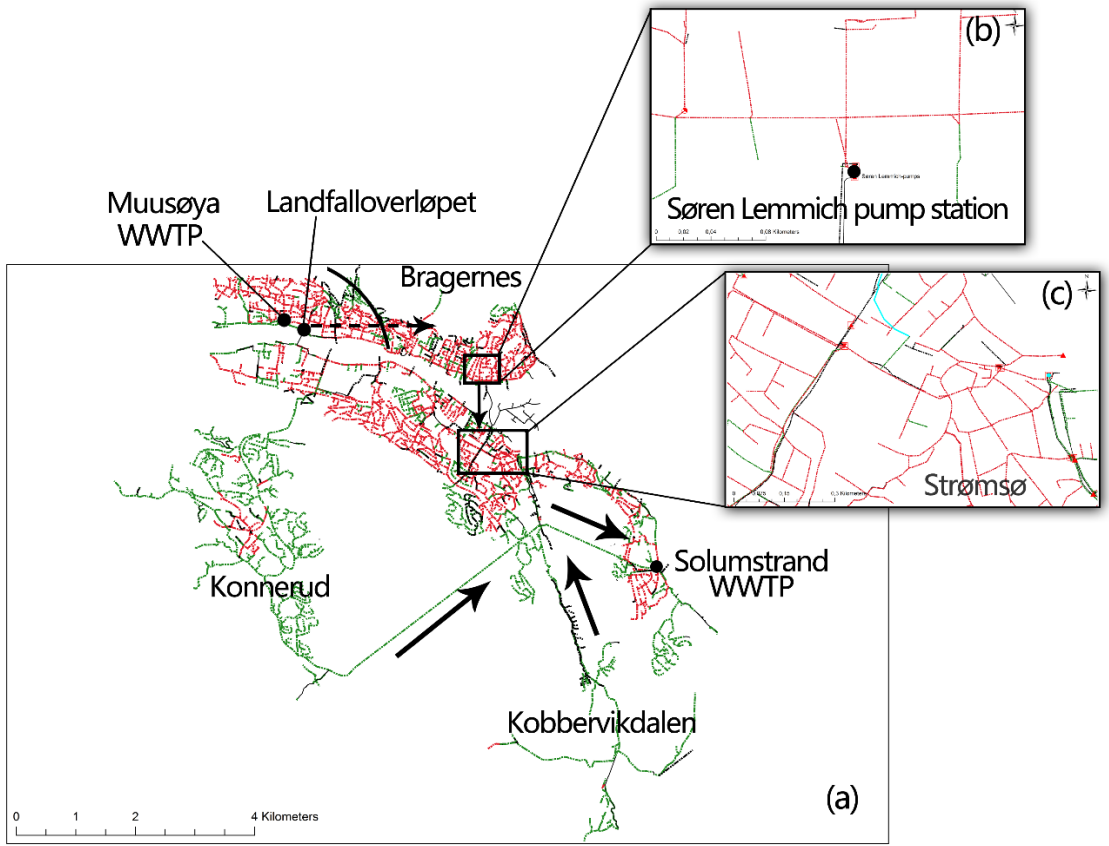

**Fig. 1.** Overview of the sewer system in Drammen

Fig.1 gives an overview of Drammen. The Drammen Fjord flow through Drammen, there are two catchments on the north of the Drammen Fjord, the Muusøya catchment and the Bragernes catchment, the curve in Fig. 1 (a) is the boundary of these two catchments. The Muusøya WWTP treats sewage collected in the Muusøya catchment. Wastewater from the Bragernes catchment is transported by the Søren Lemmich pump station (Fig. 1 (b)) from the north of the Drammen Fjord to south (the Strømsø catchment, Fig. 1 (c)). Afterward, sewage from the Bragernes catchment and the Strømsø catchment merges with wastewater from the Konnerud catchment and the Kobbervikdalen catchment, then discharge to the Solumstrand WWTP. We generalize the concept of IBWT to sewer system management, the drainage area of the Muusøya WWTP can be regarded as the 'donor basin', and the drainage area of the Solumstrand WWTP can be regarded as the 'receiving basin'. If flow to the Muusøya WWTP already exceeded its capacity and the Solumstrand WWTP still has leftover treatment capacity, part of wastewater can be conveyed to the Solumstrand WWTP for treatment. The ultimate objective of ICWT is to balance the distribution of sewer flow and uneven WWTP treatment capacities in different catchments.

Therefore, three specific questions are raised: First, whether ICWT could reduce the overflow? Second, what are the individual and combined effects of ICWT and structural measures such as storage tank? Third, it is obvious that under the ICWT scheme, the Søren Lemmich pump station will become the bottleneck of the whole system. The Søren Lemmich pump station will be requested to operate with high sensitivity, if the Søren Lemmich pump station cannot pump wastewater timely, the ICWT will only bring extra burden to the Bragernes catchment rather than mitigate overflow. The operation of a pump station highly depends on the water level information. Pumps will be activated when the water level reaches the start level of pumps. The operation of a pump station can be enhanced if accurate water level prediction information can be provided (Chiang et al., 2010). To timely operate a pump, enhance decision-making or give enough response time for operators, it is imperative to find a model that can provide the multi-step ahead water level information (Liu et al., 2016; Chang et al., 2014; Chen et al., 2014).

In present practice, the assessments of the effectiveness of nonstructural measures count on hydraulic models mainly (Autixier et al., 2014; Lucas and Sample, 2015; Chiang et al., 2010; Seggelke et al., 2005). Hydraulic models allow engineers gain insight into the functioning and effects of nonstructural measures (Chiang et al., 2010). So that the output of hydraulic models is suitable for the first and the second question. However, the implementation of hydraulic models requires perfect foreknowledge of the sewer system.

Besides, calibration, simulation and operation of hydraulic models is a time-consuming manual process. The hydraulic models can only provide information based on previous or current rainfall events. The aforementioned disadvantages of hydraulic models limit its application for question 3 (El-Din et al., 2002).

Indeed, question 3 is a hydrologic time series problem. In recent years, there is a significant rise in the number of machine learning approaches applied to hydrologic modeling and forecasting (Nourani et al., 2014). Unlike hydraulic models that derived from the hydraulic and hydrological process, the machine learning approaches learning from data without human intervention. Moreover, the trained machine learning algorithms can produce future hydrological data by being fed with current and previous data. Abovementioned advantages of machine learning have stimulated researchers to absorb it into studies about the sewer system (Yu et al., 2013; Montserrat et al., 2015; Granata et al., 2016; Zhang et al., 2016; Mounce et al., 2014).

The machine learning methods have been less active in the past decade during a period called artificial intelligence (AI) winter (Marçais & de Dreuzy 2017). In recent years, with the computer program (Google DeepMind's AlphaGo) defeated Go game world champion (Silver et al., 2016), there is renewed interest in machine-learning methods. The breakthrough technology behind AlphaGo is state of the art branch of machine learning - deep learning. The deep learning is a topic that is making big waves now, in addition to AlphaGo, another typical application of deep learning is the latest Google translation system. The new Google translation system vastly improved the translation quality, brought service nearly to the level of human translators (Google, 2016). The game changer behind the latest Google translation system is a kind of Recurrent Neural Network (RNN), Long Short Term Memory (LSTM) (Hochreiter and Schmidhuber 1997).

When performing translation, the model has to consider not only the current word, but also the other words in the sentence or even paragraph. Data with this kind of context information called sequential data. Time series data are the most popular form of sequential data, stimulated by the success of LSTM on machine translation, a few studies have explored the power of LSTM on traffic time series forecasting (Hsu, 2017).

In a case study using speed data from a sensor in Beijing, China, Ma et al., (2015) made a comparison between LSTM and RNN, SVM and traditional time series models, and LSTM outperforms other methods on time forecasting. In Tokyo, Song et al., (2016) developed an LSTM based system for predicting human mobility and transportation mode at a citywide level. Using traffic time series data from a highway around Oslo, Kanestrøm (2017) compared LSTM with recent advances of Stacked Sparse Auto Encoder (SSAE) and Deep Neural Network (DNN) on time series forecasting. The results found that the LSTM model always outperformed other models. Although LSTM has shown its superior performance, to the best of the author's knowledge, there are no prior reports about the application of LSTM in the urban hydrology studies, the effectiveness of LSTM need to be investigated.

The objective of this study has three components: (1) assessing the feasibility of ICWT, (2) studying the individual and combined effects of ICWT and storage tank, and (3) evaluating the performance LSTM on hydrologic time series prediction.

## 2. Method and materials

*2.1 The Regnbyge.no sewer monitoring system*

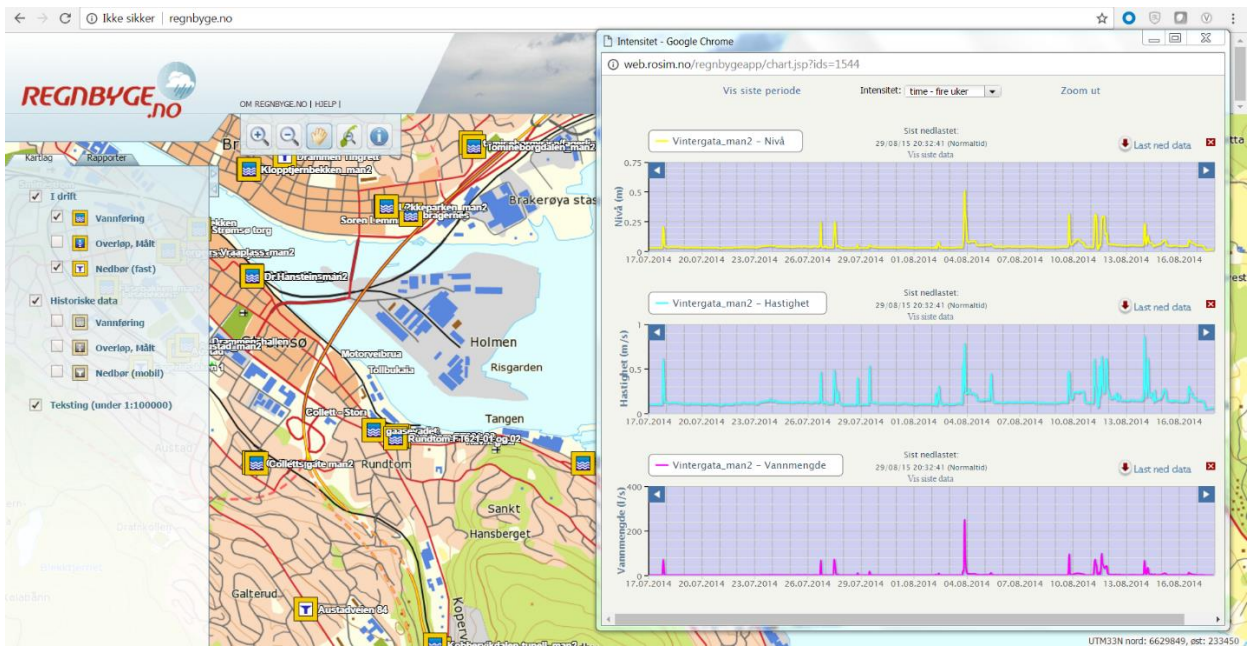

**Fig. 2.** The user interface of the Regnbyge.no sewer monitoring system

For the purpose of monitoring the sewer system as well as collecting data for model development, the Rosim AS, Norway developed a sewer monitoring system, Regnbyge.no, at the initial phase of the Regnbyge 3M project. The Regnbyge.no system consists of a number of water level sensors, velocity sensors from NIVUS GmbH, Germany and rain gauges deployed in Drammen. These sensors and rain gauges transmit collected data wirelessly to the data center at Rosim AS. A spatial database is employed to ease the process of searching, editing and managing of the collected data. Furthermore, a web Geographic Information System (GIS) is used to visualize data in the user interface of the Regnbyge.no system. Fig. 2 is the screenshot of the Regnbyge.no sewer monitoring system.

## 2.2 *Hydraulic model*

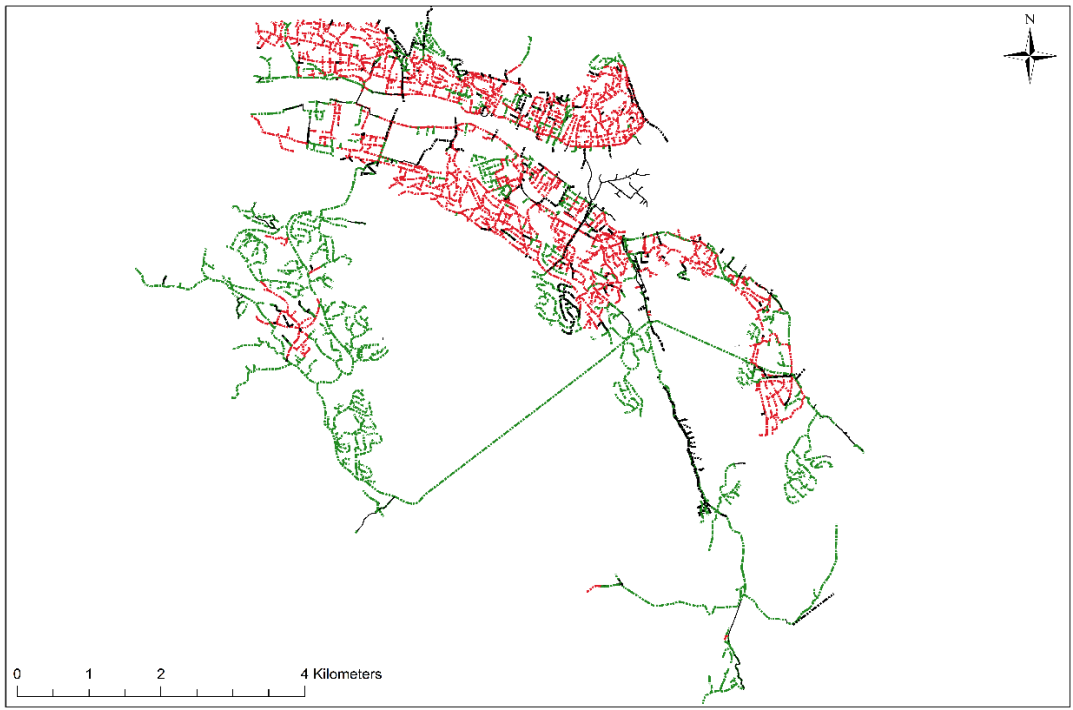

**Fig. 3.** Hydraulic model for the sewer system of Drammen

To test whether could ICWT reduce overflow, as well as study the individual and combined effects of ICWT and storage tank in considering the whole system behavior (question 1 and 2), a fully detailed hydraulic model was developed. Fig. 3 shows the hydraulic model for the sewer system of Drammen, the model consists of 9113 pipes, 9094 manholes, 129 weirs, 78 pumps and 39 outlets. The software used for hydraulic model development is Rosie, which is an ArcGIS extension developed by Rosim AS. The Rosie software maintains the interface and all the functions of ArcGIS, while using the MOUSE DHI as the computational engine. Interest readers may refer to the website of Rosie (http://web.rosim.no/index.php/tjenester/modellering-av-vann-og-avlopsnett/, in Norwegian) for more details.

In the present study, the direct response from the rainfall is calculated by the time–area (T-A) curve method A. The Rainfall Dependent Inflow/Infiltration (RDII) model is used to calculate the runoff generated from the previous hydrological processes. The pipe hydrodynamic computation is based on Saint-Venant continuity and momentum equations. The MOUSE RTC (Real Time Control) module is used to simulate different control strategies.

## 2.3 Machine learning

To provide the multi-step ahead water level prediction for managers to make decisions about pump operation. The performance of different machine learning methods, e.g. traditional algorithms such as Support Vector Regression (SVR), Feed-forward Neural Network (FFNN), traditional RNN and recent advances in deep learning (LSTM and Gated Recurrent Unit (GRU)) are evaluated.

Support Vector Machine (SVM) methods such as SVR was the major competitor of the neural network family. Although neural networks such as deep learning are nowadays dominating artificial intelligence technology, however, once upon a time, neural networks were almost unnoticed as they were overshadowed by the SVM (Cortes & Vapnik, 1995). SVR is a subcategory of SVM designed for regression problems. SVR is a kind of linear model, but it could solve nonlinear problems by using a kernel to transfer data into a feature space, and then use a linear learning mechanism to learn a nonlinear function.

FFNN is one of the most classical neural network architectures, which is comprised of input layer, hidden layer, and output layer. There are some neurons in each layer and different layers are connected by weights and bias. The FFNN first computes the weighted sum of the inputs, which can be mathematically represented as:

$$s = \sum_{i=1}^{n} w_i x_i + b \tag{1}$$

Where $w_i$ represents the weights, $x_i$ is the inputs, $b$ is the bias. Afterwards, the computed weighted sum $s$ is fed into the neuron. The neuron uses an activation function to transfer the weighted sum $s$ into the output. Usually, the FFNN are trained by using Back propagation (BP) method, BP defines the relative importance of weights for input to a neuron use chain rule of differentiation, through adjusting the weights, the FFNN reduces differences between observed and predicted values.

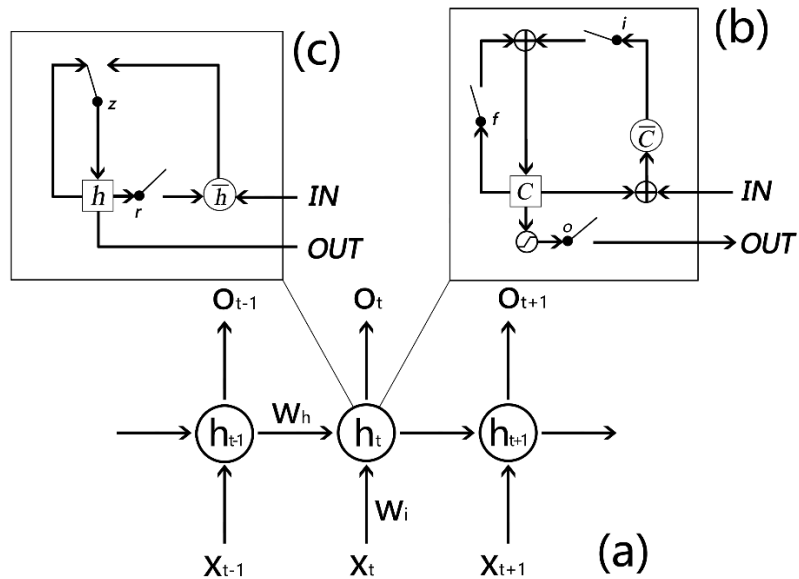

**Fig. 4.** Schematic of RNN, LSTM and GRU

The memory function of human brain inspired the concept of RNN (Elman, 1990). In addition to the weighted sum of input values, RNN also takes the state of the hidden neuron at the previous time steps as input for the next time step. As shown in Fig. 4 (a), the hidden neuron output at time step t is calculated by the equation:

$$h_t = f(w_h\ h_{t-1} + w_i\ x_t + b) \tag{3}$$

Where $h_t$ is state of the hidden neuron at the time step t, $h_{t-1}$ is state of the hidden neuron at the time step t-1, $w_i$ and $w_h$ are weights between input values and hidden neurons, and between hidden neurons respectively, $f()$ is the activation function.

Vanishing and exploding gradients, i.e. the partial derivative calculated by the chain rule of differentiation going through the network either get very small and vanish, or get very large and explode, are common problems in the training of the RNN. When train the RNN use BP (also known as backpropagation through time (BPTT)), the chain rule of differentiation not only along the direction of hidden layer and weights, but also along each time steps. Because the error of derivation accumulates through time steps, it will be extremely hard to learn and tune the parameters of the earlier layers.

LSTM was invented to combat with vanishing and exploding gradients problem. Different from traditional RNN, the LSTM uses a memory cell and three gates to control information in the hidden neuron, with on/off of the gates, information can get into, stays in or read from the cell.

Fig. 4 (b) shows the neuron in the hidden layer of LSTM, i, f and o represent the input, forget and output gate respectively. c and $\bar{c}$ denote the memory cell and the new memory cell. The principal of the memory cell in LSTM can be mathematically represented by the following equations:

Input gate:

$$i_t = \sigma_g(W_i * x_t + U_i * h_{t-1} + V_i°c_{t-1} + b_i) \tag{12}$$

Forget gate:

$$f_t = \sigma_g(W_f * x_t + U_f * h_{t-1} + V_f°c_{t-1} + b_f) \quad (13)$$

Output gate:

$$o_t = \sigma_g(W_o * x_t + U_o * h_{t-1} + V_o°c_{t-1} + b_o) \quad (14)$$

Cell state:

$$c_t = f_t°c_{t-1} + i_t°\overline{c_t} \quad (15)$$

$$\overline{c_t} = \sigma_c(W_c * x_t + U_c * h_{t-1} + b_c) \quad (16)$$

Output vector:

$$h_t = o_t°\sigma_h(c_t) \quad (17)$$

Where $x_t$ is the input vector. $W$, $U$, $V$, and $b$ are parameters for weights and bias. ° represents the scalar product of two vectors, $\sigma_g$ is the sigmoid function, $\sigma_h$ and $\sigma_c$ are the hyperbolic tangent function, for a given input z, the output of the hyperbolic tangent function is:

$$f(z) = \frac{e^z - e^{-z}}{e^z + e^{-z}} \quad (18)$$

The GRU is a recent advance in neural networks (Cho et al., 2014; Chung et al., 2014). As a variant of LSTM, the GRU also use a gating mechanism to learn long-term dependencies but its structure is much more simplified compare with LSTM. Fig. 4 (c) shows the gating mechanism of GRU. GRU has only a reset gate and an update gate. The GRU combines the input and forget gates into an update gate to balance between previous activation and the candidate activation. The activation of h at time t depends on h at the previous time and the candidate h (the $\overline{h}$ in Fig. 4 (c)). The update gate z decides how much of the previous memory to keep around. The GRU unit forgets the previously computed state when the reset gate r off.

The GRU is formulated as:

$$z_t = \sigma_g(W_z * x_t + U_z * h_{t-1} + b_z) \quad (19)$$

$$r_t = \sigma_g(W_r * x_t + U_r * h_{t-1} + b_r) \tag{20}$$

$$h_t = z_t °h_{t-1} + (1 - z_t)°\overline{h_t} \tag{21}$$

$$\overline{h_t} = \sigma_h(W_h * x_t + U_t * (r_t°h_{t-1}) + b_h) \tag{22}$$

Where $x_t$ is the input vector, $h_t$ is the output vector, $z_t$ is the update gate vector, $h_t$ is the reset gate vector. $W$, $U$ and $b$ are parameters for weights and bias. ° represents the scalar product of two vectors, $\sigma(.)$ is the sigmoid function. $\sigma_g$ represent the sigmoid activation function, $\sigma_h$ represent the hyperbolic tangent activation function.

In this study, the LSTM, GRU, RNN and FFNN is implemented using Keras. Keras is a Python based high-level deep learning library. It is running on top of TensorFlow or Theano. TensorFlow is used as the backend of Keras in this study. TensorFlow is an open-source deep learning software released by Google in 2015. The SVR is implemented using Python machine learning library Scikit-learn.

## *2.4 Model performance criteria*

The performance of different models is evaluated by three criteria, the root mean square error (RMSE), Nash-Sutcliffe Efficiency (NSE) and the $R^2$. The calculation of RMSE as shown below:

$$RMSE = \sqrt{\frac{\sum_{i=1}^{n}(Y_i^{obs} - Y_i^{sim})^2}{n}} \tag{11}$$

The NSE is calculated by the following equation:

$$NSE = 1 - \left[\frac{\sum_{i=1}^{n}(Y_i^{obs} - Y_i^{sim})^2}{\sum_{i=1}^{n}(Y_i^{obs} - Y^{mean})^2}\right] \tag{12}$$

The equation for $R^2$ is:

$$R^2 = \left[\frac{(\sum_{i=1}^{n}(Y_i^{sim} - Y_{sim}^{mean})(Y_i^{obs} - Y^{mean}))^2}{\sum_{i=1}^{n}(Y_i^{sim} - Y_{sim}^{mean})^2 \sum_{i=1}^{n}(Y_i^{obs} - Y^{mean})^2}\right] \tag{13}$$

In the three above listed equations:

$Y_i^{obs}$ = the $i$-th observed data.

$Y_i^{sim}$ = the $i$-th simulated data.

$Y^{mean}$ = mean value of observed data.

$Y_{sim}^{mean}$ = mean value of simualted data.

$n$ = number of observed data.

# 3. Results

## *3.1 Calibration of the hydraulic model*

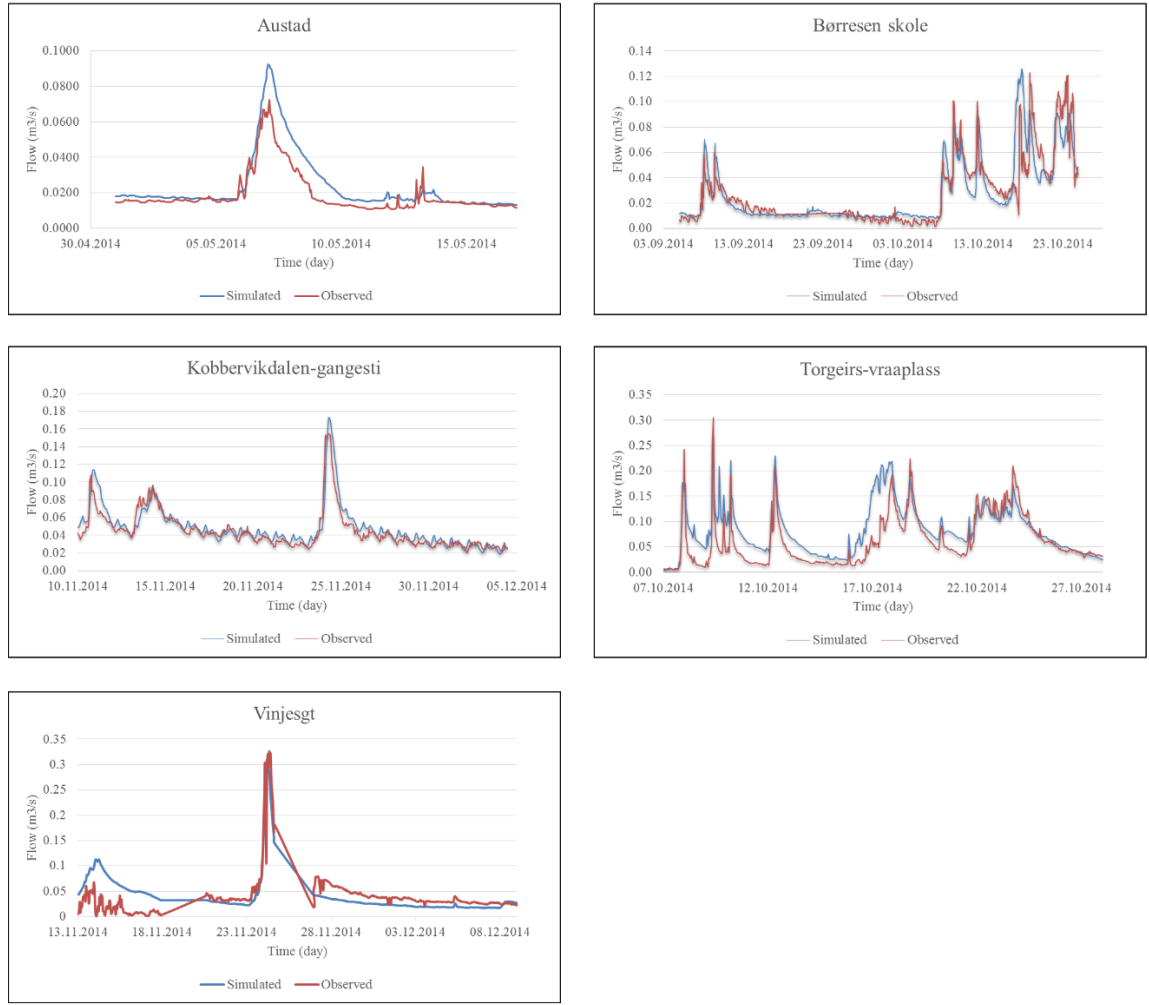

**Fig.5.** Hydrographs of observed versus hydraulic model simulated flow

**Table 1.** Calibration results of hydraulic model

| Monitoring site | NSE | $R^2$ | RMSE |
|---|---|---|---|
| Austad | 0.54 | 0.91 | 0.008 |
| Børresen skole | 0.68 | 0.71 | 0.014 |
| Kobbervikdalen-gangsti | 0.80 | 0.85 | 0.010 |
| Torgeir-vraaplass | 0.51 | 0.67 | 0.036 |
| Vinjesgt | 0.53 | 0.57 | 0.029 |

Flow data recorded by the regnbyge.no sewer monitoring system is used to calibrate the hydraulic model. Fig. 5 shows the hydrographs of the hydraulic model outputs versus the recorded values at five monitoring sites. It clearly indicates that the simulated values are consistent with the recorded values. Table 1 lists the model performance criteria. All the criteria show acceptable values. Results display in Fig. 5 and Table 1 confirmed a high reliability of the hydraulic. The calibrated model is used in the following scenario simulations.

## *3.2 Scenario simulations*

In the current phase of the Regnbyge 3M project, in order to mitigate overflow from the Muusøya WWTP, the Drammen municipality is planning to construct a storage tank at Landfalloverløpet, however, due to dense buildings and population, the maximum size of the storage tank is restricted to 20,000 m$^3$, which is insufficient to deal with the current overflow situation. The proposed ICWT solution is expected to compensate insufficient capacity of the storage tank.

Eight scenarios are designed to study individual and combined effects of the storage tank and ICWT on overflow mitigation. The operation of the storage tank and ICWT is simulated using the RTC module in Rosie. When the inflow to the Muusøya WWTP exceeds its maximum capacity, the wastewater is diverted to the storage tank, if the storage tank is full, then the ICWT is activated to convey wastewater to the Brageners catchment. Table 2 gives descriptions of the eight scenarios. The annual long simulation was run continuously from January 01, 2014 to December 31, 2014 to simulate sewer system behaviors.

**Table 2.** Descriptions of designed scenarios for hydraulic simulation

| Scenario | Descriptions |
|---|---|
| Scenario 1 | Baseline scenario, without any overflow control measures |
| Scenario 2 | Only construct a 1,000 m$^3$ storage tank at Landfalloverløpet |
| Scenario 3 | Only construct a 5,000 m$^3$ storage tank at Landfalloverløpet |
| Scenario 4 | Only construct a 20,000 m$^3$ storage tank at Landfalloverløpet |

| Scenario 5 | Only implement the ICWT between the Muusøya catchment and the Brageners catchment |
| Scenario 6 | 1,000 m³ storage tank + ICWT |
| Scenario 7 | 5,000 m³ storage tank + ICWT |
| Scenario 8 | 20,000 m³ storage tank + ICWT |

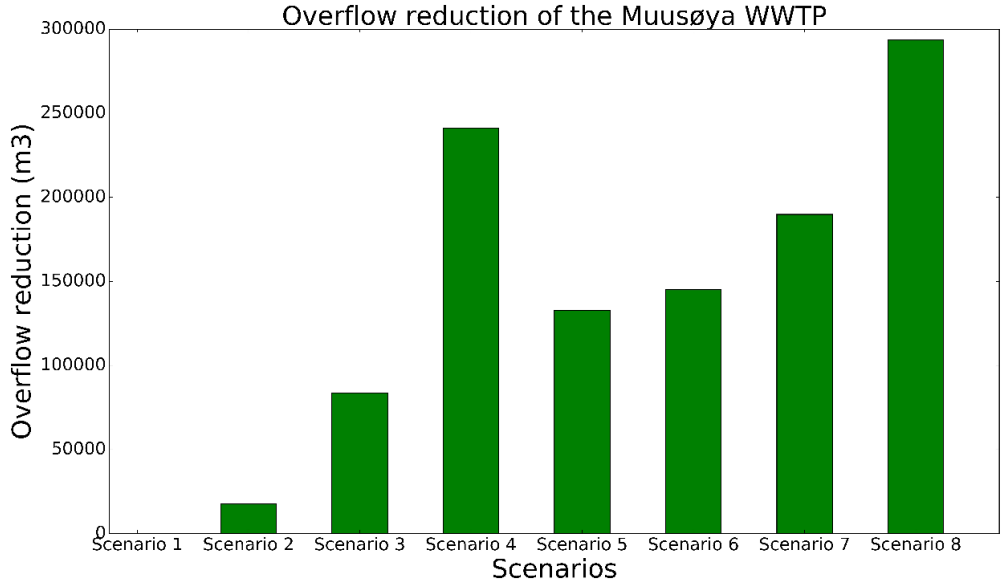

**Fig 6.** Volume of reduced overflow from the Muusøya WWTP

Fig. 6 displays overflow reduction from the Muusøya WWTP under different scenarios. The overflow from the Muusøya WWTP is 456,171 m$^3$ for scenario 1. There is a clear tradeoff between storage tank sizes and overflow reduction. For scenario 2 with the smallest storage tank size, the overflow reduction is only 17,576 m$^3$. A 5000 m$^3$ storage tank (scenario 3) reduces 83,768 m$^3$ overflow. The largest storage tank (scenario 4) decreased the overflow more than 50% compare to scenario 1. Compare to ICWT, the storage tank seems less efficient. Scenario 5 with only ICWT reduces more overflow than scenario 3. With same storage tank size, the application of ICWT substantially enhances overflow reduction. The volume of overflow reduction for scenario 2, scenario 3 and scenario 4 are increased from 17,576 m$^3$, 83,768 m$^3$, 241,112 m$^3$ to 145,012 m$^3$, 189,801 m$^3$, 293,449 m$^3$, respectively for scenario 6, scenario 7 and scenario 8.

**Table 3.** Total overflow of the sewer system

| Scenario | Total overflow volume (m$^3$) |
|---|---|
| Scenario 1 | 2,096,668 |
| Scenario 2 | 2,071,522 |
| Scenario 3 | 2,001,256 |
| Scenario 4 | 1,837,773 |
| Scenario 5 | 1,976,164 |
| Scenario 6 | 1,954,927 |
| Scenario 7 | 1,895,657 |
| Scenario 8 | 1,771,521 |

Alongside with the overflow from the Muusøya WWTP, the total overflow are also analyzed. Total overflow volume for scenario 1 is approximately 2,096,668 m$^3$. Scenario 2 only decreases the overflow volume by 1.20%, which is almost not distinct to scenario 1. With the maximum available storage tank dimension (20,000 m$^3$), the total overflow reduced to 1,837,773 m$^3$, a reduction of 12.35%. Only implement ICWT reduces total overflow to 1,976,164 m$^3$, which is more efficient than a 5,000 m$^3$ storage tank. Compare to scenario 3 and scenario 4, the ICWT further decrease total overflow volume from 2,071,522 m$^3$ and 2,001,256 m$^3$ to 1,954,927 m$^3$ and 1,895,657 m$^3$. Scenario 8 has the lowest overflow volume with both maximum storage tank size and ICWT.

**Table 4.** Volume of overflow from the Søren Lemmich pump station

| Scenario | Overflow from the Søren Lemmich pump station (m$^3$) |
|---|---|
| Scenario 1 | 32,940 |
| Scenario 5 | 52,076 |
| Scenario 6 | 45,079 |
| Scenario 7 | 34,862 |
| Scenario 8 | 18,650 |

One obvious concern about implementing ICWT is whether it will bring extra burden to the Søren Lemmich pump station. Volume of overflow from the Søren Lemmich pump station are investigated. For scenario 5, the total amount of overflow is greater than scenario 1, it means only implement ICWT bring extra overflow to the Søren Lemmich pump station, although the total overflow reduced. Applying the storage tank resulted in a reduction in overflow from the Søren Lemmich pump station. For scenario 6, a 1,000 m$^3$ storage tank reduces overflow compare to scenario 5 but the overflow still higher than scenario 1. Overflow volume for scenario 7 is similar to scenario 1, it means a 5,000 m$^3$ storage tank can relieve the burden of the Søren Lemmich pump station brought by ICWT. For scenario 8, one can observe that total overflow, overflow from the Muusøya WWTP and overflow from the Søren Lemmich pump station are reduced. It can be concluded that ICWT is an efficient overflow mitigating measure, however, in considering the whole system behavior, a storage tank with the size range from 5,000 m$^3$ to 20,000 m$^3$ is suggested to implement in conjunction with ICWT.

## 3.3 Machine learning

The hydraulic simulations clearly demonstrate the viability of ICWT (question 1 and 2). The purpose of this section is to explore the potentiality of machine learning, particularly deep learning in hydrological time series forecasting (question 3). To train the forecasting models, water level data of the Søren Lemmich pump station and corresponding rainfall data collected by the Regnbyge.no system is used. The data are collected from March 20, 2014 to December 1, 2014, there are 73,597 records with a temporal resolution of 5 min. To validate the generalization of the machine learning algorithm, the data are divided into two subsets: training set and testing set. Data from the first 75% were used for training, and the remaining 25% were used for testing. Table 5 gives the summary statistics of the datasets. In the present study, data are scaled to the range [0, 1] before training. After developing the models, the scaled values are rescaled to real values.

**Table 5.** Summary statistics of the water level data

| Model stage | Max water level (m) | Average water level (m) | Standard deviation of water level | Max rainfall (mm/s) | Average rainfall (mm/s) | Standard deviation of rainfall |
|---|---|---|---|---|---|---|
| Training | 9.51 | 2.77 | 1.44 | 10.15 | 2.62 | 2.11 |
| Testing | 9.18 | 3.17 | 2.15 | 9.07 | 2.68 | 2.08 |

Considering technical details of the pump operation, the outputs of the models are a lead-time up to 24 steps (2 hours). The inputs of the models are selected by applying cross-correlation and autocorrelation to the datasets, using the XCORR and AUTOCORR function in MATLAB R2016a. Afterward, the LSTM is implemented through trial and error experiments, different hyper-parameters such as the number of hidden layer, number of hidden neuron, and different optimizers (RMSprop, Adadelta, Adam, Adamax) are tried. The optimal structure of LSTM has two hidden layers with 128 hidden neurons in each layer, the Adam is chosen as the optimizer. Additionally, Dropout is used to prevent overfitting. During training, Dropout (Hinton et al., 2012) temporary discard part of neurons from the neural network. This procedure can be regarded as generating a number of "thinned" neural networks during training but use a single un-thinned neural network in testing (Srivastava et al., 2014). A dropout ratio of 0.35 is selected in this study. The LSTM is trained for 200 epochs with a batch size of 128. After ascertaining the structure of LSTM, the performance of LSTM is compared with other models. To make a fair comparison, the GRU, FFNN and RNN remain the same structure with LSTM. The performance of SVR is subject to the kernel function and the other parameters such as gamma, C and epsilon, the grid search method in the Python Scikit-learn library is used to find the optimal kernel and parameters for SVR. The optimized SVR used in this study is an RBF kernel SVR with a gamma value of 0.5, a C value of 5 and an epsilon value of 0.01.

**Table 6.** Summary of model performance

| Lead time | Performance criteria | Deep learning methods | | Traditional methods | | SVM |
|---|---|---|---|---|---|---|
| | | GRU | LSTM | RNN | FFNN | SVR |
| 20 minutes (4 steps) | $R^2$ | 0.8969 | 0.9014 | 0.9011 | 0.9006 | 0.8674 |
| | RMSE | 0.6987 | 0.6790 | 0.6774 | 0.6799 | 0.7886 |
| | NSE | 0.8943 | 0.9002 | 0.9006 | 0.8999 | 0.8653 |
| 40 minutes (8 steps) | $R^2$ | 0.8942 | 0.8965 | 0.8738 | 0.8631 | 0.8573 |
| | RMSE | 0.7040 | 0.6929 | 0.7438 | 0.7175 | 0.8178 |
| | NSE | 0.8927 | 0.8961 | 0.8628 | 0.8586 | 0.8552 |
| 1 hour | $R^2$ | 0.8900 | 0.8926 | 0.8376 | 0.8169 | 0.7902 |
| (12 steps) | RMSE | 0.7186 | 0.7091 | 0.7991 | 0.8377 | 0.8959 |
|  | NSE | 0.8882 | 0.8912 | 0.8417 | 0.8022 | 0.8174 |
|  | $R^2$ | 0.8819 | 0.8832 | 0.8298 | 0.8072 | 0.7942 |
| 1 hour 20 minutes (16 steps) | RMSE | 0.7435 | 0.7553 | 0.8021 | 0.8518 | 0.9042 |
|  | NSE | 0.8803 | 0.8765 | 0.8143 | 0.8143 | 0.8038 |
|  | $R^2$ | 0.8721 | 0.8707 | 0.7903 | 0.7643 | 0.7750 |
| 1 hour and 40 minutes (20 steps) | RMSE | 0.7856 | 0.7740 | 0.8572 | 0.9258 | 0.9199 |
|  | NSE | 0.8664 | 0.8703 | 0.7993 | 0.7524 | 0.7669 |
|  | $R^2$ | 0.8620 | 0.8670 | 0.7756 | 0.7538 | 0.7390 |
| 2 hours (24 steps) | RMSE | 0.8107 | 0.7961 | 0.8972 | 0.9232 | 0.9541 |
|  | NSE | 0.8578 | 0.8628 | 0.7854 | 0.7155 | 0.7330 |

The results for the multi-step-ahead water level forecasting within the test period are provided in Table 6. Overall, the LSTM outperforms other models based on $R^2$, RMSE and NSE criteria. The performance of GRU is slightly worse than the LSTM but the difference is marginal. LSTM and GRU have higher $R^2$ and NSE values than the other three models for long term predictions. As expected, longer time step caused less accuracy. The models perform consistently well for 4-step and 8-step ahead forecasting, whereas as the forecasting lead time exceeds 1 hour (12 steps), significant differences appear among their performances. The RMSE of the models increases, while the $R^2$ and NSE decrease, as the forecasting step increases.

The value of the three criteria indicates that the LSTM has the ability to predict water level with long lead time. The performance statistics of the LSTM yields an $R^2$ of 0.8707 and 0.8670 respectively, an RMSE of 0.7740 and 0.7961 respectively, and an NSE of 0.8703 and 0.8628 respectively for 20-step and 24-step ahead water level prediction. The results prove that with the gating mechanism and memory cell, the LSTM can substantially improve the accuracy of multi-step-ahead water level forecasts.

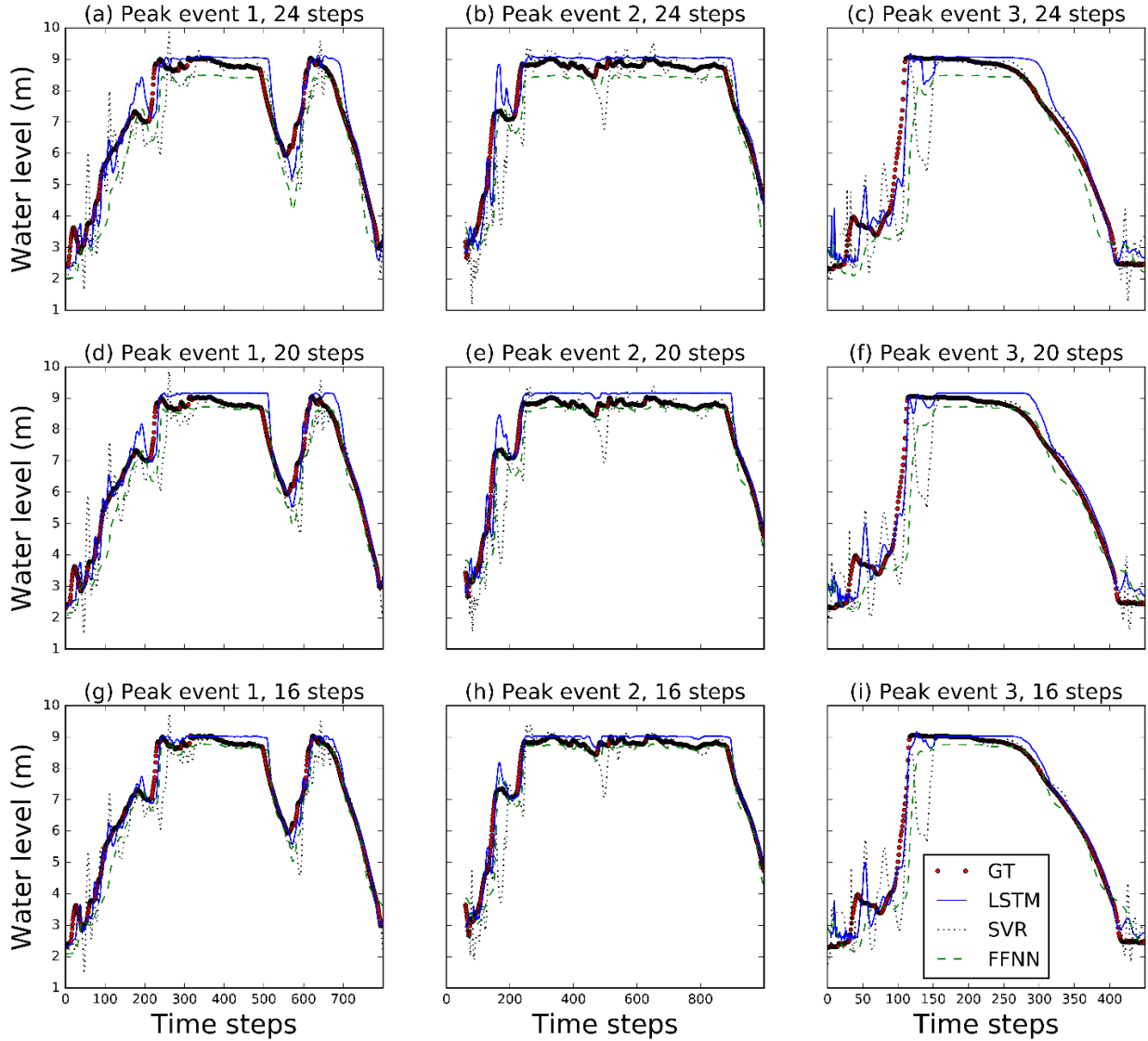

**Fig. 7.** Hydrographs of observed versus 24, 20 and 16-step ahead forecasted water levels of the LSTM, FFNN and SVR from three peak events

To intuitively illustration the model performance, the hydrographs of Ground Truth (GT) value, LSTM, FFNN and SVR are presented in Fig.7. Fig.7 is observed versus 24, 20 and 16-step ahead forecasted water levels of the three selected models from three peak events. In general, LSTM is still superior to others. Fig.7 shows that the LSTM predicted data are generally consistent with the observed data. It can be seen from Fig.7 that the developed LSTM model is able to predict the water level despite significant variations in water levels during rainfall events. In Fig.7 (a), (c), (d) and (f), one can observe that FFNN have significant time-lag phenomena, whereas LSTM significantly mitigates this problem, it can be observed that water level increases are anticipated timely by LSTM. Besides, the LSTM is able to capture the major trends and peaks of observations. While FFNN often under-estimates the peak value, as shown in Fig.7 (a)-(c). Compare to LSTM and FFNN, SVR presents strong fluctuations at the rising limbs and has sudden drops at some peaks.

## 4. Conclusion

This paper delineates a novel ICWT solution for mitigating sewer overflow with minimal construction works. Hydraulic model is developed to test the effectiveness of ICWT and study extra burden received by the Søren Lemmich pump station. To further enhance the operation of the Søren Lemmich pump station, a representative deep learning technology, LSTM, is employed to provide multi-step ahead water level predictions. Several useful findings can be concluded from this study.

1) Most previous studies about sewer overflow control only focus on a single component of the sewer system. To control the overflow in a systematical way, the sewer system should be reconsidered as a whole system to let individual sewer components cooperate in a holistic way. As indicated by hydraulic simulations of eight scenarios, the ICWT could efficiently reduce total overflow from the sewer system and overflow from the Muusøya WWTP, however, the ICWT may bring extra burden to the Søren Lemmich pump station. In considering the whole system behavior, the ICWT is suggested to act in concert with the storage tank.

2) Nonstructural overflow mitigating solutions usually request key facilities to be operated with high sensitivity. The risks of overflow may be increased if accurate hydrological time series prediction information cannot be provided. Five different machine learning models, including deep learning methods (LSTM and GRU), the traditional RNN and FFNN, and the SVR, are compared in this study. Experiments demonstrated that the LSTM is superior to other methods. The LSTM model is capable of forecasting multi-step ahead hydrological time series, and therefore can be a great tool for sewer system managers.

3) Studies relative to sewer systems require the modeling of complex and dynamical urban hydrological processes. There are two approaches of models existed, in the extensively used hydraulic model approach, hydrological/hydraulic principals are explicitly modeled. However, the complicated model construction, calibration and computation make hydraulic models less adequate for real time purpose. On the other hand, the implicit machine learning approach could provide predictions in real time, but it cannot describe the hydrological/hydraulic behavior of sewer system in detail. To solve practical problems, engineers or researchers should consider combining the advantages of both approaches to let them complement each other.

4) The potential power of deep learning is fascinating, ubiquitous sensors are monitoring infrastructures such as sewer system, and collecting a large amount of data. The fusion of sensors, actuators and algorithms will become the backbone of future cities. Data analytics is the sticking point for leveraging intelligent infrastructure management. Given the revolutionary strides made by deep learning in recent years, there are many prospective interests of deep learning for hydrological studies.

## Acknowledgements

This work has been supported by the Regnbyge-3M project (grant number 234974), which is granted by the Oslofjord Regional Research Fund. The authors would like to thank the engineers from Rosim Company for their supports.

## Conflict of Interest

None